# AI Design, Design AI, Human-Centred AI and the Theatre of the Absurd
– the language, life and times of a UX designer

**Author:** Rebekah Rousi; rebekah.rousi@uwasa.fi
**Affiliation:** School of Marketing and Communication, University of Vaasa

**Introduction**
Repetition, meaninglessness, empty signifiers, uncertainty and chaos seem to be the spirit of the times – times in which technology and design are set to solve the unsolvable. Amidst the despair of global warming, warfare and political unrest, the international population is worried about what robots and artificial intelligence (AI) will mean in terms of future work, and humanity in general as great efforts are being placed on creating thinking and emotional machines. Designers and design have been laden with the task of embracing AI and its potential through design thinking, presenting on stages around the world about the empowerment AI affords in terms of matters such as data-intensive human-centred design, and relieving humans of the jobs that no one wants to do, to name two. It is not just the design of AI systems that is generating hype, but also the prospect of the potential of AI to either perform or enhance creative tasks such as design itself. AI is being emphasised as a tool of the designer[1], not a means of replacing the designer, yet developers and corporate still cannot resist efforts to raise AI as the designer of the future. Corporate examples of the strive towards AI as fashion designer for instance, can be seen in the developments of Amazon, Stitchfix and Myntra to name some[2].

When thinking of Glenn Parson's[3] description of design as a deliberate and intentional act, it seems quite absurd to understand machines (AI) as designers or directors of the creative process. Moreover, for decades now designer intentionality and additionally design stance, are integral to understanding and recognising design as a cognitive action and conscious agency[4] towards shaping the lived human world (and all it's contents) through modifying and augmenting its structure[5]. So, is there any risk of AI taking over the designer role, if it does not have its own consciousness and lived experience?

This argument is supported in part by user experience (UX) designer Nick Dauchot (2018) who poses questions on the nature of creativity and empathy and their possibilities within AI technology in relation to the designer's work. Dauchot concludes that AIs possess creativity only to the extent of their programmers, and that these technologies are only an exhibit of the programmers' capacities and imaginations. To combine the idea that design is an act of consciousness, it can be posited that in addition to empathy, creativity requires intentionality[6] – it is intentionality (intention) that defines a product (service, system, object or expression) as creative. Without the intention, or human associative understanding (rationality) of the matter that is being generated, one is left simply with a product of the Chinese Room[7]. This means that the consciousness[8] and conscious experience of the designer should be present while producing creative work. It is at this point – consciousness and intentionality in the production of design – that one may look at the UX designer: human-centred, at the hub of AI innovation and highly involved in picking up the pieces.

---

[1] See e.g., Algorithm-driven design - How artificial intelligence is changing design, Available at: https://algorithms.design/; or Harshani Chathurika's (2019), 'Future of artificial intelligence in design'; and Jess Holbrook's 'Designing human-centered AI products' from Google Design for instance.

[2] From a practical industrial perspective, Vishal Kalia (2018) uses examples to discuss the possibilities of AI in replacing the roles of fashion designers in his Youtube video, "Will Artificial Intelligence Replace Fashion Designers? [Amazon, Stitch Fix, Myntra]". https://www.youtube.com/watch?v=5BMjo1SLXKQ; David Vandegrift delves deeper into exploring the possibilities of creative AI from a philosophical perspective in "Can Artificial Intelligence be Creative?" https://medium.com/@DavidVandegrift/can-artificial-intelligence-be-creative-40e7eac56e71. Vandegrift ponders over whether or not machines will be more intelligent than humans, and actually argues for the eventuality of creative AI. Whether or not his argument stands, depend on one's definition of creativity – particularly relating to creativity as expression, versus creativity as thought.

[3] Glenn Parsons, *The philosophy of design* (Hoboken, NJ: John Wiley & Sons, 2015).

[4] Nathan Crilly, "The design stance in user-system interaction," *Design Issues 27*, no. 4 (2011): 16-29.

[5] Daniel C Dennett, "Intentional systems," *The Journal of Philosophy 68*, no. 4 (1971 ): 87-106; "The Interpretation of Texts, People and Other Artifacts," *Philosophy and Phenomenological Research 50* (1990):177-94

[6] John R. Searle, "The intentionality of intention and action," *Inquiry 22*, no. 1-4 (1979): 253-280; John R. Searle, "The Chinese room revisited," *Behavioral and brain sciences 5*, no. 2 (1982): 345-348.

[7] The Chinese room described later in the text refers to an experiment in which subjects (non-Chinese speaking) were asked to behaviorally respond to messages written in Chinese through relying on a reference book. In this, subjects more or less mimicked the symbol processing of computers, whereby a 'programmer' creates the correct responses in the reference book that are retrieved via 'commands' written in the Chinese texts.

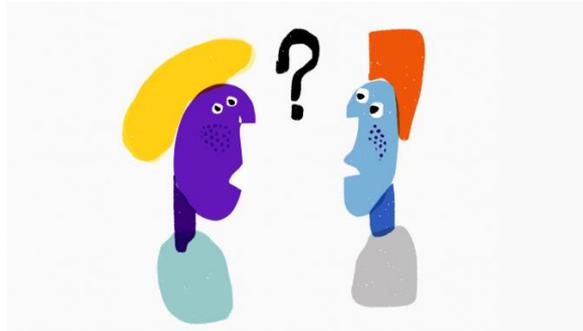

Fig. 1 Number #2 of "9½ plausible and/or absurd UX industry predictions for 2018"[9]

Caught in the midst of the AI design hype is a broad and diverse group of professionals known as user experience (UX) designers. This professional group comprises individuals spanning a range of educational and experiential backgrounds, from software engineering, computer science, psychology, cognitive science to design, art, languages, ethnography, philosophy and chemistry to name some. While they themselves, understand that they cannot indeed design experience, they do try to instil the understanding that they can, in fact, influence the design process in such a way that certain types of qualities and emotional reactions can/should be experienced while interacting with a product or service[10]. The key element behind the UX movement is the acknowledgement that: a) humans are diverse[11]; b) their experience matters[12]; and c) in order to be able to design successfully, designers need to understand the users and their experiences while using products, services and systems. The reason behind such a broad knowledge community as in the case of UX designers, is the awareness of context and use purpose combined with the user-design relationship. Thus, "not all UX jobs are the same"[13], and in fact, the cases, objectives and approaches may be equally as diverse as the team of professionals labelled as UX designers.

Adobe developer Nick Babich describes the UX designer phenomenon succinctly when he states, "...its function still remains a mystery to many... it is not always immediately clear what they actually do day-to-day."[14] Activities that UX designers engage in may entirely involve coding, or perhaps may be completely human-focused and/or conceptually based. From Babich's perspective, UX design can be broken down into six main tasks: 1) product research, 2) creating personas and scenarios, 3) information architecture, 4) creating wireframes, 5) prototyping, and 6) product testing. Yet, it is known that these tasks vary from company to company depending on the product and service offering, business model and organisation of the business. Furthermore, UX design may be more user/customer research intensive (i.e., anthropology based), co-design rich, and/or even organisational strategy-focused. Already from this perspective pinpointing the role of intentionality, whose intentionality and what intentionality within the design process may seem baffling, even before the addition of AI. So, did you hear the news about Bobby Watson?

---

[8] David J. Chalmers, "Facing up to the problem of consciousness," *Journal of Consciousness Studies 2*, no. 3 (1995): 200-219.
[9] Craig Phillips, "9 ½ Plausible and/or Absurd UX Industry Predictions for 2018," UX Collective (2017). https://uxdesign.cc/9-%C2%BD-plausible-and-or-absurd-ux-industry-predictions-for-2018-542897a6446a
[10] Nick Babich, "What does a UX designer actually do?" (2017). https://theblog.adobe.com/what-does-a-ux-designer-actually-do/
[11] Marc Hassenzahl and Noam Tractinsky, "User experience-a research agenda," *Behaviour & information technology 25*, no. 2 (2006): 91-97.
[12] Donald A. Norman, *Design of everyday things: Revised and expanded* (New York: Hachette, 2013).
[13] Barcelona Code School, "What background and skills do I need to become a UX designer?" (2019). https://barcelonacodeschool.com/what-background-and-skills-do-I-need-to-become-a-UX-designer
[14] Babich, "What does a UX designer actually do?"

> **MR. SMITH:** Which Bobby Watson do you mean?
> **MRS. SMITH:** Why, Bobby Watson, the son of old Bobby Watson, the late Bobby Watson's other uncle.
> **MR. SMITH:** No, it's not that one, it's someone else. It's Bobby Watson, the con of old Bobby Watson, the late Bobby Watson's aunt.
> **MRS. SMITH:** Are you referring to Bobby Watson the commercial traveler?
> **MR. SMITH:** All the Bobby Watsons are commercial travelers.
> **MRS. SMITH:** What a difficult trade! However, they do well at it.[15]

These days, all the Bobby Watsons are UX Designers – and they claim to do well at it. As Craig Phillips[16] from the UX Collective notes, "A recruiter buddy of mine told us about a new role coming up that's right up your alley. UX Innovation in Hiring Manager Designer (Developer). Main roles include UX writing, design, and development of innovative job titles and descriptions. Candidate must have 5-30 years experience managing a confused team coping with existential crises and dark thoughts." This job description and professional field as a whole seems apt for the task of developing the mammoth sky ship(s) known to many as AI, or cognitive computing, as often times the concept of AI itself is as fuzzy as the teams developing it. For this reason, the next section describes Absurdism in the context of 20th century theatre and its key philosophical underpinnings. The paper then connects absurdism to UX designers and their role in AI design – particularly human-centred AI – discourse through observations made from the blogs of UX designers working for major AI corporate players (IBM, Google, Adobe etc.).

**Absurdism**
*The Theatre of the Absurd (ToA)*, or *Absurdism* was a brief and fairly unofficial genre of theatre that emerged during the 1950s and early 60s[17]. With a small number of playwrights, works which arose within this genre could be distinguished by an illogical progression (or non-progession) of narrative, uncertainty, meaninglessness and elements that when combined generate an atmosphere which is anti-aesthetic (dis-harmonic) by nature. Some of the work, such as witnessed in Samuel Wey Beckett's *Waiting for Godot*[18], presents repetition without firm continuity, despair through familiarity yet strangeness, and a painful interaction between the audiences' lived experience of time and sanity, with the above mentioned characteristics of the plays. Others including Eugene Ionesco's *The Bald Soprano*, highlight the unintentional, intentionless relationship between language, its communication from one person to another, and lived experience.

Written while Ionesco attempted to learn English, there was an aim to capture the pointlessness and meaninglessness of human existence as presented through people's conversations in the English language (as well as English itself). The language presented relationships and countless networks of people, in addition to bogus narratives and anecdotes, rendering the activities, recollections and social circumstances that humans often consider valuable, as mere babble – without a beginning, or an end, or even a point. The technique Ionesco embraced was influenced by his experiences with the Assimil method – a process of listening to phrases and emptily repeating them without thought for meaning. This mode of language learning echoes with the process and results of John Searle's[19] famous Chinese Room experiment in which non-Chinese speaking participants sat in a room, equipped with a reference book. They were then handed messages in Chinese characters. The participants needed to utilise the reference book to search for the correct responses to the characters they had received in the letter. Participants were able to successfully respond to these Chinese letters via the use of the Chinese symbol book without any understanding of the translation into English. Thus, dialogue and interaction were able to occur on the level of code, symbols and representation,

---

[15] Eugene Ionesco, "The Bald Soprano," *The Bald Soprano and other plays,* First printed in 1958 (New York City, NY: Grove Press, 2015), 4.
[16] Phillips, "9 ½ Plausible and/or Absurd UX Industry Predictions for 2018"
[17] Martin Esslin, *The theatre of the absurd* (New York City, NY: Vintage, 2009).
[18] Samuel Beckett, *Waiting for Godot* (London: Faber and Faber, 1988 [1956]).
[19] Searle, "The intentionality of intention and action"; Searle, "The Chinese room revisited"

without any conscious understanding or intentionality on behalf of at least one (the participants) of the parties.

This form of *empty* language use and expression – symbolic, yet lacking any real intention or meaning – is characteristic of the ToA. Upon viewing the first ever showing of *Waiting for Godot*, a prisoner of San Quentin described his understanding of the play in which the protagonist of the play (Godot) never appears, as representing society[20]. For after sitting through the painstaking developments, or lack thereof, of the performance, the audience was still left waiting, unsure of what or who Godot was. Yet, they somehow understood that if he/she/it actually emerged, he/she/it would be a disappointment.

The ToA interweaves with trends put forth by numerous cultural movements present in literature and fine art, seen in works of Cubism, Surrealism, abstractionism, and by Franz Kafka, as well as James Joyce. Ionesco defined the term "absurd" as "that which is devoid of purpose"[21]. Moreover, in reference to Kafka, Ionesco goes on to describe how once humans are severed from their metaphysical, transcendental and religious roots, they become lost in a metaphysical anguish and frustration. Throughout plays presented under the banner of ToA are characteristics of the "senselessness of life... devaluation of ideals, purity and purpose"[22]. Rather than succinctly argue about the absurd nature of humanity through distinctive literary and theatrical quality, adhering to the most intricate of conventions, ToA enacts it. The absurdity spans from language to relationships, and how language constructs or disrupts the purity of human-to-human connections. Interestingly, Beckett's study of Proust demonstrates many of the mechanisms present in his own future work, including the illusion of friendship, the inability to possess loved ones, attempts to communicate where communication is impossible. Beckett stated, that "There is no communication because there are no vehicles of communication"[23].

Ironically, not only UX designers, but often times entire companies and business models fail to communicate, as the vehicle through which they are communicating, and the vehicle they are also trying to communicate seem absent. With the rhetoric of tech biz (technology business) spinning around from 'design' to 'UX' to 'AI' it is no wonder that the industry and general public find themselves surrounded by empty signifiers[24] – signs, symbols and language that have lost significance due to repetition and breadth of use. Moreover, similarly to the risks of being trapped in the discourse of one's own profession and its trends, Becket maintained that there were dangers of operating in one's own native language as individuals become stuck in and blind to the logic of these languages – vulnerable to automatic acceptance of meanings and associations without reflection of what the languages are actually constructing. This could be considered a loss of intentionality when communicating in one's native language, contra to the pitfalls presented in Searle's Chinese room experiment.

**Absurdism meets UX meets AI design**
The lack of intentionality and then justification (intentionality) behind creating purpose for unintentional thought replication systems within discourse of AI, and especially Design AI, establishes a humorous discourse between both the public and the technology – not to mention the *designer*. What are they thinking? Moreover, what are the machines thinking? What do they intend to achieve, solve or provoke in their latest designs? During times when UX designers are not engaged in their day-to-day tasks of conducting user research, validating and testing concepts, prototypes and products with users, designing, selling and writing UX copy,[25] they are also reaching out to the broader public. This public outreach that is often in the form of blogs or vlogs, most frequently deals with defining UX and their operations as professionals. Moreover, UX designers as a group of professionals are highly involved in leading the design AI movement as the people who present, represent and describe the human component AI, generally known as human-centred AI. Furthermore, rather than providing the form for AI within the products and services themselves, UX designers are assembling a face, through narrative and linguistic constructions (public speaking, guides etc.) of how AI an unfold, be used and core components that could and/or should be considered.

---

[20] Martin Esslin, *Absurd Drama* (London, UK: Penguin, 1965).
[21] Esslin, *The theatre of the absurd*, 23.
[22] ibid., 24.
[23] Samuel Beckett, *Proust* (New York City, NY: Grove, 1931/1989): 47
[24] Ernesto Laclau, "Why do empty signifiers matter to politics," *Emancipation(s)* (London: Verso, 1996): *36*-46.
[25] Suji Niroshan, "What do I (UX designer) actually do?" (2019). https://uxdesign.cc/who-is-a-ux-designer-b484855e32b1

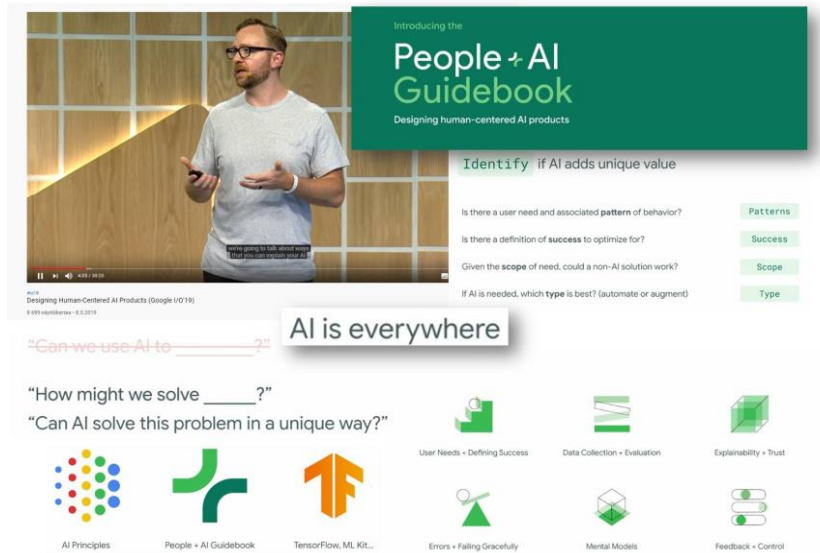
Fig. 2 Designing human-centered AI products[26]

These talks, observable at almost every modern tech, business and design event, and accessible through the likes of Youtube and TED channels, can be viewed as ToA productions. UX designers are playing out and enacting the language of an entity that almost no one knows anything about (Godot). By the end of the presentations the audience has acquired a new language based on the frameworks and guidelines set forth by the presenter, yet, they are still waiting and not exactly knowing any more about AI than when they first began listening. One recent talk by Google designer Jess Holbrook[27] demonstrates this form of theatre, whereby, AI is given a body through guidelines, which are in themselves embodied in the talks of UX designers. Holbrook's presentation (see Figure 2) introduces the audience to a human-centered AI guidebook that assists designers in six main areas: 1) user needs – defining success; 2) data collection – evaluation; 3) explainability + trust; 4) errors + failing gracefully; 5) mental models; and 6) feedback and control. These areas do not differ greatly to traditional guidelines (heuristics) of usability, what is amusing however, is that it rests neatly alongside two other guidebooks for designers on AI – AI Principles and TensorFlow, ML Kit...

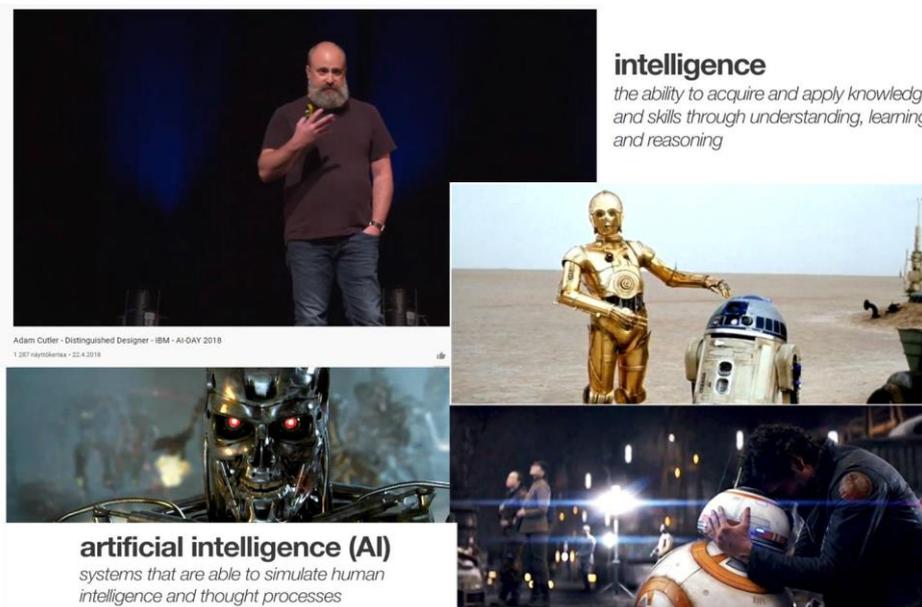
Fig. 3 Adam Cutler – Distinguished Designer, IBM AI Day 2018

---

[26] Jess Holbrook, "Designing human-centered AI products," Google I/O'19.
https://www.youtube.com/watch?v=rf83vRxLWFQ
[27] ibid.

Seemingly concrete, tangible and informative, Adam Cutler[28] from IBM takes a different approach to giving AI a body. Cutler states directly that designing "for something that is still lacking form or understanding is kind of a challenging thing." Most of the images, representations and subsequent meanings constructed by Cutler and his team for both the general public and even back-end professionals are derived from popular culture (see Figure 3). These not only give AI a look and feel, but they also come attached to complex sets of emotions and expectations. Thus, conceptual constructions of AI play with, from and off the public imagination, are laden with human qualities, as well as associations of unrealistic utopia or despair, destruction and post-humanism (and/or post-humanity). One may pause to wonder why corporate and designers are in fact investing so much into what could ultimately be considered the end of humankind.

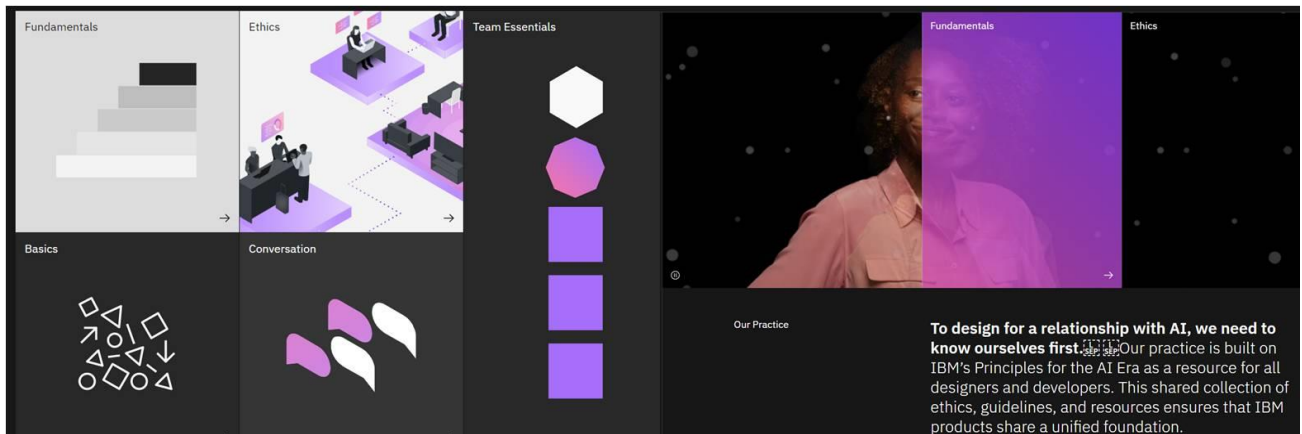

Fig. 4 IBM Design for AI webpages

To counter this, IBM has also adopted a seemingly socially responsible approach to enacting their AI (see figure 4), whereby its AI webspace presents the technology in simple, biteable chunks – as if the company is fully aware and in control of its developments. Then, there are the human aspects of communication (conversion), team essentials (tool kit) and on-trend ethics. The ethics component sings to the drama acknowledged and unfolding in Cutler's presentation, highlighting that more needs to be understood about ourselves as humans in order to design human-AI relationships and once again referring to guidelines (principles) as a foundation of practice and development.

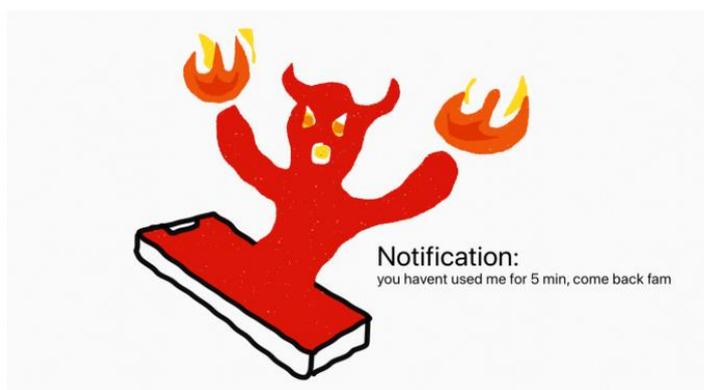

Fig. 5 Number #8 of "9½ plausible and/or absurd UX industry predictions for 2018"[29]

---

[28] Adam Cutler, "Adam Cutler - Distinguished Designer," IBM AI Day 2018. https://www.youtube.com/watch?v=GrbSdUYswX8
[29] Phillips, "9 ½ Plausible and/or Absurd UX Industry Predictions for 2018"

Yet, more than anything this attests to the absurdity of Phillips'[30] number 8 of the 9½ Plausible and/or Absurd UX Industry Predictions for 2018 (see figure 5), whereby "Troubling products will proliferate, and we'll all just get used to it." Here, Phillips describes how 2018 would possibly be the beginning of the dystopian future whereby UX designers focus on "'killin it' while 'hands down' creating some 'habit-forming' 'experiences' for the homogeneous mass of 'users' whom, btw, we have 'empathized' with"[31]. Phillips goes on to explain that even though designers may consider themselves ethical and empathic, in their projects, designers successfully separate their emotion from their tasks through reasoning that: 1) there were orders from management; 2) economic necessity; and 3) life and work systems are complex and will always be filled with ethical and moral controversy. What is more, as Holbrook[32] states, AI should simply be seen as a tool and, "At the end of the day, tools are just enablers."

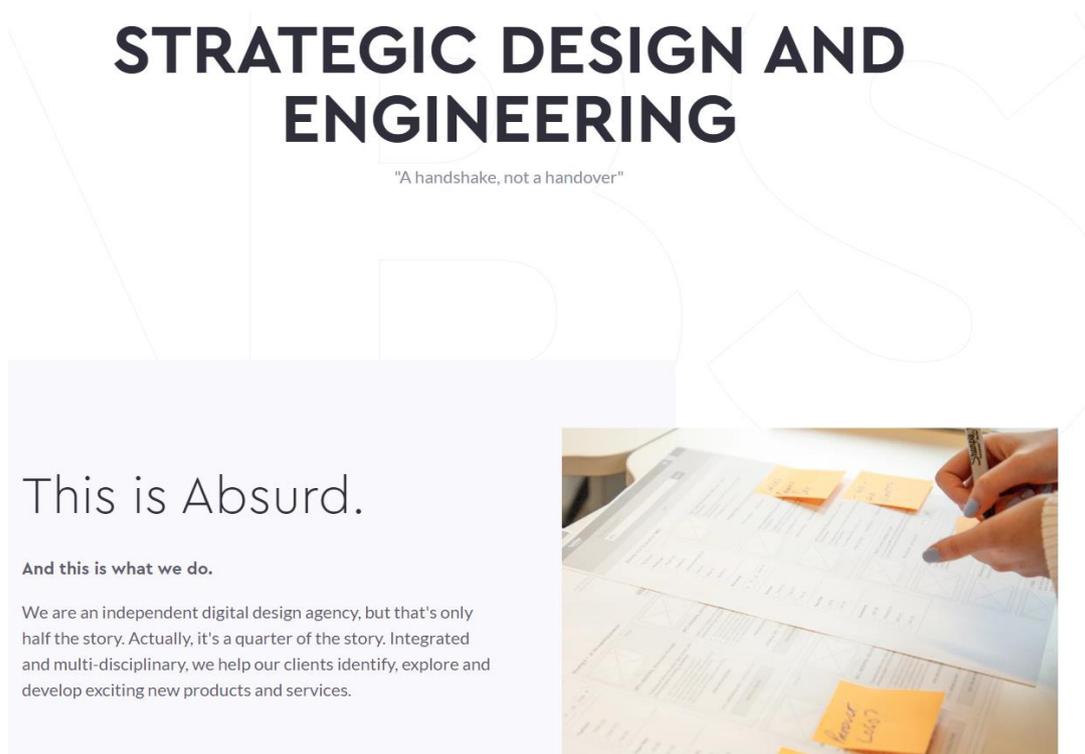

Fig. 6 This is Absurd – strategic design and engineering.

**Conclusion meets inconclusion**
Returning to Esslin[33] and ToA, it may be observed that ToA attacks the norms and certainties of political and religious hegemony. The theatre genre aimed to dismay and disrupt complacency amongst its audiences, and to throw the harsh facts of humanity in the face of those watching. Currently, the address of environmental (e.g., Greta Thunberg), economic (e.g., Stephane Garelli) and social issues (e.g., Michael Moore) to name some at all business and tech forums aims to do this too. However, where ToA attempted to deconstruct these institutions at their grassroots, social (including ethical and moral), economic and environmental responsibility are the spectacle of disruption on stage. That is, in this society of the spectacle[34] it may not be surprising if these seemingly *sustainable* and ethical directions may, in the end, unfold as a series of empty signifiers – the smoke masks on top of the smoke masks. The meaning and purpose becomes meaningless if, at the end of the day, the messages have been produced to no effect – no genuine intentionality, just another message. ToA challenges its audiences to see humanity for what it is – complex, absurd, mysterious and meaningless (contrary to the Theatre of the Techbiz, *ToT*). The human condition is painful, in the end illusions dissolve and there are no effective easy solutions[35]. As the name of a strategic design and engineering company suggests (see figure 6), "This is Absurd".

---

[30] ibid.
[31] To be overly explicit, remember Samuel Becket's thoughts on Proust?
[32] Jess Holbrook, "Designing human-centered AI products," n.p.
[33] Esslin, *Absurd Drama*.
[34] Guy Debord, *Society of the spectacle* (Canberra: Hobgoblin, 2002).

The progression of AI design discourse has occurred over numerous decades. Repetition of insight, speculation and development makes each new technological iteration and its drawbacks another scene of *Waiting for Godot*. The idea that AI adopts roles, not simply of mechanical, repetitive factory workers, but also of lovers[36] and creative practitioners. Or that it is considered a 'tool' for design practice, when considered from the stand-point of intelligence, may incite desperation in the human condition. From the perspective of user experience, AI as a tool is interesting[i]. User experience by nature intends to tap into the very essence of humanity - human thought and experience. And/or should AI be thought of more in terms of augmented intelligence (or augmented cognition) that merely extends the designers intentionality through enabling the previous impossible in terms of human information processing capabilities (Clark & Chalmers, 1995; Schmorrow, 2005), or autonomous intelligence (Holbrook, 2019)? In all cases, ToA, UX design and AI, at the end of the day we are simply left with a performance of language, false beliefs (and possible values), illusions and a desperate form of empathy.

---

[35] Esslin, *Absurd Drama*
[36] Rebekah Rousi (2018) wrote about the prospects of robot emotions and genuine mutual love between humans and robots in, "Me, my bot and his other (robot) woman?"

---

[i] See for example, https://www.awwwards.com/AI-driven-design